\documentclass[review,longtitle, authoryear]{elsarticle}
\usepackage{graphicx}
\usepackage{amssymb}
\usepackage{alltt}
\usepackage{subfigure}

\usepackage[labelsep=space]{caption}
\usepackage{amsmath}
\newcommand{\degree}{\ensuremath{^\circ}}

\newcommand{\eg}{e.g., }

\newcommand{\fig}[1]{Fig.\ \ref{f:#1}}

\newcommand{\figs}[2]{Figs.\ \ref{f:#1}--\ref{f:#2}}
\newcommand{\Fig}[1]{Figure \ref{f:#1}}

\newcommand{\tbl}[1]{Table \ref{t:#1}}

\newcommand{\code}[1]{\texttt{#1}} % for fixed-width font

\newcommand{\figcap}[1]{\item{\bfseries Figure \ref{f:#1}. }}

\usepackage[left, pagewise, displaymath, mathlines]{lineno}

\begin{document}
\title{Numerical Simulations of Collisional Disruption of Rotating Gravitational Aggregates: Dependence on Material Properties}
  \author[umd]{R.-L.~Ballouz\corref{cor1}}
  \ead{rballouz@astro.umd.edu}
  \author[umd]{D.C.~Richardson}  
  \author[oca]{P.~Michel}
  \author[oca,umd]{S.R.~Schwartz}
  \author[tu,umd]{Y.~Yu}

  \cortext[cor1]{Corresponding author}

  \address[umd]{Department of Astronomy, University of Maryland, College Park MD 20740-2421, United States}
  \address[oca]{Lagrange Laboratory, University of Nice Sophia Antipolis, CNRS, Observatoire de la C\^{o}te d'Azur, C.S. 34229, 06304 Nice Cedex 4, France}
  \address[tu]{School of Aerospace, Tsinghua University, Beijing, 100084, Republic of China}

\begin{abstract}
Our knowledge of the strengths of small bodies in the Solar System is limited by our poor understanding of their internal structures, and this, in turn, clouds our understanding of the formation and evolution of these bodies. Observations of the rotational states of asteroids whose diameters are larger than a few hundreds of meters have revealed that they are dominated by gravity and that most are unlikely to be monoliths; however, there is a wide range of plausible internal structures. Numerical and analytical studies of shape and spin limits of gravitational aggregates and their collisional evolution show a strong dependence on shear strength. In order to study this effect, we carry out a systematic exploration of the dependence of collision outcomes on dissipation and friction parameters of the material components making up the bodies. We simulate the catastrophic disruption (leading to the largest remnant retaining 50\% of the original mass) of km-size asteroids modeled as gravitational aggregates using \code{pkdgrav}, a cosmology $N$-body code adapted to collisional problems and recently enhanced with a new soft-sphere collision algorithm that includes more realistic contact forces. We find that for a range of three different materials, higher friction and dissipation values increase the catastrophic disruption threshold by about half a magnitude. Furthermore, we find that pre-impact rotation systematically increases mass loss on average, regardless of the target's internal configuration. Our results have important implications for the efficiency of planet formation via planetesimal growth, and also more generally to estimate the impact energy threshold for catastrophic disruption, as this generally has only been evaluated for non-spinning bodies without detailed consideration of material properties. 
\end{abstract}
\maketitle

\section{Introduction}
\indent Collisions dominate the formation and evolution of small Solar System bodies (SSSBs). In the early stages of the Solar System, planetesimals interacted with one another in a dynamically cold disk \citep[see][]{Levison10}. This allowed planet-size objects to form through collisional growth. Later, asteroid families formed through the catastrophic disruption of parent bodies. Outcomes of collisions between SSSBs are divided into two regimes: those dominated by material strength and those dominated by self-gravity \citep{Holsapple94}. Since the dominant source of confining pressure for planetesimal-size SSSBs is self-gravity rather than material strength, they can be assumed to be gravitational aggregates \citep{Richardson02}. Hence, the collisions can often be treated as impacts between rubble piles, the outcomes of which are dictated by collisional dissipation parameters and gravity \citep{Leinhardt00, KorycanskyAsphaug09}. Understanding the effects that contribute to changes in the mass (accretion or erosion) of gravitational aggregates is important for collisional evolution models of the early Solar System \citep[e.g.,][]{LeinhardtRichardson05, Weidenschilling11}. The outcomes of impacts in these models are characterized by a catastrophic disruption threshold $Q^{\star}_{D}$ \citep[e.g.,][]{BenzAsphaug99}, which is the specific impact energy required to disperse permanently half the total mass of the system, such that the largest remnant retains the other half of the system mass.\\
\indent The specific impact energy at which a body disrupts catastrophically is dependent on the ``strengths'' of the body's material. These are the tensile, compressive, and shear strengths. There is evidence that small asteroids with sizes above a few hundred meters are likely to be cohesionless, and, therefore, lacking tensile strength \citep{Pravec08}. However, these bodies are not completely strengthless. While the dominant confining pressure is self-gravity, their granular nature gives them the capability to withstand considerable shear stress when under pressure. The shapes of the components physically impede their neighbors from flowing around them. The macroscopic effect is a pile of granular material with an angle of repose that is characteristic of that material. The angle of repose depends not only on the bulk shapes of the components, but on the other material properties that dictate the frictional forces to which they are subject. A stack of perfectly smooth, frictionless cannonballs can be stable simply because of their rigid shapes (so long as the bottom plane is fixed). Friction is not required to maintain a non-zero angle of repose, however friction can increase this angle \citep{Zhou02,Richardson12}. The envelope of permissible equilibrium shapes of a granular body is typically parameterized by a Mohr-Coulomb angle of internal friction, $\phi$. The angle $\phi$ varies from $0\degree$ to $90\degree$, where a body with $\phi=0\degree$ is a fluid, and higher values represent materials that are able to resist higher shear stresses. Normal terrestrial granular materials have $\phi\sim30\degree$ to $40\degree$. Very little is known about the internal structure of SSSBs, despite the increasing amount of observational data from ground- and space-based resources \citep[e.g.,][]{Belton94,Veverka00}. The most direct way to measure the strength of material is by physically breaking it. However, attempts to damage SSSBs \citep[e.g.,][]{Ahearn05} are expensive and, therefore, limited in number.\\
\indent In the past decade, advances in analytical and numerical studies have been made that attempt to correlate shape and spin states of SSSBs with their possible internal configurations. \citet{Holsapple01} determined lower limits for $\phi$ for various C-, S-, and M-type asteroids based on their shapes and spins. \citet{Walsh08,Walsh12} and \citet{Holsapple10} studied the shape and spin changes of self-gravitating bodies in response to YORP-induced increases in their angular momentum. \citet{Walsh12}, using spherical particles, found that $\phi$ is also influenced by the size distribution of the particles that make up the body. These studies have provided the necessary groundwork to begin to understand the internal make-up and actual strength of SSSBs.\\
\indent \citet{KorycanskyAsphaug09} studied binary collisions of rubble piles modeled as collections of polyhedral particles. Using a non-penetrating-rigid-body approach, they studied the mass loss outcomes for three different dissipation parameters and two different size distributions of particles (a monodisperse size distribution and a polydisperse power-law size distribution with a power-law index of $-1$). They found that both of these factors affect the catastrophic disruption threshold, with $Q^{\star}_{D}$ increasing for higher dissipation and for a power-law size distribution of particles. For constant mass, a power law distribution of particles would have a larger internal surface area than a monodisperse distribution. The larger number of particle contacts would allow more collisional energy to be dissipated through friciton and inelastic collisions. However, their study used a single power law index; therefore, it is uncertain whether this result is true for any power law distribution of particles. Furthermore, their work used a limited number of particles ($N\sim10^3$) and modeled dissipation using arbitrary friction and restitution parameters, and they did not study the effect of rotation on collision outcome.\\
\indent \citet{Ballouz14} found that catastrophic disruption is sensitive to the initial pre-impact rotation of the target. Since rotational evolution depends on the internal structure of SSSBs, it is unclear whether rotationally enhanced collisional mass loss also depends on the material properties of a small body. \\
\indent In this work, we study the dependence of catastrophic disruption outcomes on the material properties of both the target and impactor, which are obtained from comparisons with laboratory experiments. Furthermore, we study whether this sensitivity could be dependent on the rotational properties of the colliding bodies. By doing so, we begin to map out the relation between the strength of a small body and collision outcomes. This will help inform planetesimal formation and evolution studies by providing physically realistic descriptions of the energies required for catastrophic disruption. This aids in delineating the transition from accretion to erosion in collision outcomes.\\
\indent We solve numerically the outcomes of rubble-pile collisions using a combination of a soft-sphere discrete element method (SSDEM) collisional code \citep{Schwartz12} and a numerical gravity solver, \code{pkdgrav} \citep{Richardson00, Stadel01}, which is needed to model the reaccumulation stage accurately. The SSDEM code allows us, for the first time, to model multi-contact and multi-frictional forces accurately as well. We compare our results to the dissipation-dependent catastrophic disruption study by \citet{KorycanskyAsphaug09}, and to the mass-ratio-dependent study by \cite{LeinhardtStewart09}.\\
\indent In Section 2 we explain the computational methods and outline the parameter space that we explore. In Section 3 we provide our results. In Section 4 we summarize and offer perspectives.
 
\section{Methodology}

We use \code{pkdgrav}, a parallel $N$-body gravity tree code \citep{Stadel01} adapted for particle collisions \citep{Richardson00,Richardson09,Richardson11}.  Originally collisions in \code{pkdgrav} were treated as idealized single-point-of-contact impacts between rigid spheres.  A soft-sphere option was added recently \citep{Schwartz12}; with this option, particle contacts can last many timesteps, with reaction forces dependent on the degree of overlap (a proxy for surface deformation) and contact history. This allows us to model multi-contact and frictional forces. The code uses a 2nd-order leapfrog integrator to solve the equations of motion, with accelerations due to gravity and contact forces recomputed each step.\\
\indent The spring/dash-pot model used in \code{pkdgrav}'s soft-sphere implementation is described fully in \citet{Schwartz12}.  Briefly, a (spherical) particle overlapping with a neighbor feels a reaction force in the normal and tangential directions determined by spring constants ($k_n$, $k_t$), with optional damping and effects that impose static, rolling, and/or twisting friction. User-defined normal and tangential coefficients of restitution used in hard-sphere implementations, $\varepsilon_n$ and $\varepsilon_t$, determine the plastic damping parameters ($C_n$ and $C_t$), which are required to resolve a soft-sphere collision (see Eq.\ 15 in \citet{Schwartz12}). The static, rolling, and twisting friction components are parameterized by dimensionless coefficients $\mu_s$, $\mu_r$, and $\mu_t$, respectively. For cohesionless material, the angle of repose $\phi$ is determined by a combination of frictional and shape properties.  Shape effects, arising from the sizes and geometries of the grains, can alone be important in determining the angle of repose, especially in material that exhibits nonexistent or weak friction and cohesion. Even for $\mu_s = 0$, rubble piles made of idealized spheres have a non-zero angle of repose, owing to cannonball stacking. Using spherical particles, \citet{Walsh12} were able to correlate the internal structure (particle size distribution) of such rubble piles with a value of $\phi$ by simulating their spin and shape evolution and comparing the results to the analytical theory developed in \citet{Holsapple01}.\\
\indent The numerical approach has been validated through comparison with laboratory experiments; \eg \citet{Schwartz12} demonstrated that \code{pkdgrav} correctly reproduces experiments of granular flow through cylindrical hoppers, specifically the flow rate as a function of aperture size, \citet{Schwartz13} demonstrated successful simulation of laboratory impact experiments into sintered glass beads using a cohesion model coupled with the soft-sphere code in \code{pkdgrav}, and \cite{Schwartz14} applied the code to low-speed impacts into regolith in order to test asteroid sampling mechanism design.

\subsection{Rubble-Pile Model}
Our simulations consist of two bodies with a mass ratio of $\sim 1:10$, specifically a stationary target with mass $M_{\mathrm{targ}}$ and a projectile with mass $M_{\mathrm{proj}}$ ($M_{\mathrm{proj}}$ = 0.1$M_{\mathrm{targ}}$) that impacts the target at a speed of $v_\mathrm{{imp}}$. Both the target and projectile are gravitational aggregates of many particles bound together by self-gravity. The particles themselves are indestructible and have a fixed mass and radius. \\
\indent The rubble piles are created by placing equal-sized particles randomly in a spherical cloud and allowing the cloud to collapse under its own gravity with highly inelastic particle collisions. Randomizing the internal structure of the rubble piles reduces artificial outcomes due to the crystalline structure of hexagonal close packing \citep{Leinhardt00, LeinhardtRichardson02}. We discuss the influence of this randomization of the internal configuration of the rubble pile on collision outcome in Section 2.3.\\
\indent For the simulations presented here, the target had an average radius of $R_\mathrm{{targ}} \sim 1.0$ km and bulk density of $\rho_\mathrm{{targ}} \sim  1.7$ g cm$^{-3}$. The projectile had an average radius of $R_\mathrm{{proj}} \sim 0.5$ km and bulk density of $\rho_\mathrm{{proj}}\sim 1.65$ g cm$^{-3}$. In order to determine the physical properties (size, shape, mass, angular momentum) of the post-collision remnants accurately, the rubble piles were constructed with a relatively high number of particles ($N_{\mathrm{targ}} = 10^4, N_{\mathrm{proj}} = 10^3$).\\
\indent We use km-size projectiles and targets since they lie near the transition between strength-dominated (which decreases with size) and gravity-dominated (which increases with size) bodies. Therefore, they are some of the weakest bodies in the Solar System, as they require the least amount of impact energy per unit mass to be disrupted \citep{BenzAsphaug99}. This allows us to simulate accurately the collisional dynamics using SSDEM, as the collision speeds required for catastrophic disruption (on the order of tens of of meters per second) do not exceed the sound speed of the material. Hence, there is no irreversible shock damage to the particles, which the code does not model currently.\\
\indent Since SSDEM models treat particle collisions as reactions of springs due to particle overlaps, the magnitude of the normal and tangential restoring forces are determined by the spring constants $k_n$ and $k_t \sim (2/7)k_n$. We approximate $k_n$ by setting the maximum fractional particle overlap, $x_{\mathrm{max}}$, to be $\sim 1\%$. For rubble-pile collisions, the value of $k_{n}$ can be given by:
\begin{equation} k_n \sim m \left(\frac{v_{max}}{x_{max}}\right)^{2},\end{equation}
\noindent where $m$ corresponds to the typical mass of the most energetic particles, and $v_{max}$ is the maximum expected speed in the simulation \citep{Schwartz12}. Thus, for our rubble-pile collisions with speeds $< 20$ m s$^{-1}$, $k_n \sim 4 \times 10^{11}$ kg s$^{-2}$. The initial separation of the projectile and target, $d$, for all cases was $\sim 4R_{\mathrm{targ}}$, far enough apart that initial tidal effects were negligible. In order for the post-collision system to reach a steady state, the total run-time was set to $\sim 3\times$ the dynamical time for the system, $1/\sqrt{G\rho_{targ}} \sim 2$ hours. Furthermore, a time-step $\Delta t \sim 3$ ms was chosen on the basis of the time required to sample particle overlaps adequately, for the choice of $x_{\mathrm{max}}$ and hence $k_n$ given above.

\subsection{Material Properties}
\indent For this study, we compared three different sets of soft-sphere parameters for the gravitational aggregates (\tbl{SSDEMparams}). Following \citet{Yu14}, we defined three groups of parameters that span a plausible range of material properties. Since the actual mechanical properties of gravitational aggregates are poorly constrained, these three sets are just a guess at the likely limits of the strength properties of monodisperse gravitational aggregates, in the absence of cohesion. The first group (``smooth'') consists of idealized frictionless spheres with a small amount of dissipation. This is about as close to the fluid case that a rubble pile can achieve while still exhibiting shear strength arising from the discrete nature of the particles (and the confining pressure of self-gravity). The second group (``glass beads'') is modeled after actual glass beads being used in laboratory experiments of granular avalanches \citep{Richardson12}. For this group, $\varepsilon_t$ was measured directly, and the friction coefficients, $\mu_s$ and $\mu_r$, were inferred from matching simulations to the experiments.\\
\begin{center}
\textit{\tbl{SSDEMparams}}.
\end{center}
\indent The properties of the third group (``gravel'') were found by carrying out simple avalanche experiments using roughly equal-size rocks collected from a streambed \citep{Yu14}.  A series of numerical simulations were performed to reproduce the typical behavior of the rocks by varying the soft-sphere parameters. The large values of $\mu_s$ and $\mu_r$ values reflect the irregular non-spherical shapes of the experimental particles. Cohesion would further increase the strength of the gravitational aggregates; however, we do not consider the effects of cohesion here, leaving an exploration of that wide parameter space for future studies.\\

\subsection{Collision Setup}
For each material property group, simulations were done with a range of impact speeds such that there was adequate coverage of the gravitational dispersal regime (collisions that result in a system losing $0.1$--$0.9$ times its total mass). The collision of two rubble-pile objects typically results in either net accretion, where the largest remnant has a net gain in mass compared to the mass of the target, or net erosion, where the target has lost mass. Alternatively, a collision could result in no appreciable net accretion or erosion \citep{LeinhardtStewart12}. These latter types of collisions, called hit-and-run events, typically occur for grazing impacts that have an impact parameter, $b$, that is greater than a critical impact parameter $b_{\mathrm{crit}}$ \citep{Asphaug10}, where $b_{\mathrm{crit}} \equiv R_\mathrm{{targ}}/(R_{\mathrm{proj}}+R_\mathrm{{targ}})$. In this paper, we focus on the dispersive regime, where the impact speed, $v_{\mathrm{imp}}$, is greater than the escape speed from the surface of the target, $v_{\mathrm{esc}}$ (assuming no rotation). The impact speeds in our simulation range from $\sim$ 5 -- 20 $v_{\mathrm{esc}}$, where $v_{\mathrm{esc}} \sim 1$ m s$^{-1}$ is the escape speed from a spherical object with mass $M_{\mathrm{tot}}=M_{\mathrm{proj}}+M_\mathrm{{targ}}$ and density $\rho = 1 $ g/cm$^{3}$. The amount of mass loss at the end of a simulation is found by measuring the final mass of the largest remnant and all material gravitationally bound to it (material with instantaneous orbital energy $< 0$) and subtracting this from $M_{\mathrm{tot}}$.\\ 
\indent Since the level of collisional dissipation is influenced by the internal configuration of a rubble pile, we expect there to be some variation in collision outcome due to the randomization arising from how the initial conditions are generated (Section 2.1). In order to characterize this uncertainty, we generated 5 randomized rubble piles to compare with our nominal gravitational aggregate (Section 3.1).\\
\indent \citet{Ballouz14} found that catastrophic disruption is sensitive to pre-impact rotation, and they constructed a semi-analytic description for the outcomes of head-on equatorial collisions (collisions where the projectile's trajectory is directed at the center of the target and is perpendicular to the target's rotation axis). However, the study was performed for a single set of SSDEM parameters that were similar to the mildly dissipative ``glass beads'' case. Therefore, in order to study whether the decrease in the catastrophic disruption threshold due to rotation is sensitive to material properties, we compare head-on collisions with no pre-impact rotation to cases where the target rubble pile has an initial spin period $P_{\mathrm{spin}}$ of 6 hours for the 3 material property groups given in \tbl{SSDEMparams}.

\section{Results}
We show graphically the mass loss outcomes of our simulations in \figs{mesc_smooth}{mesc_gravel}. We plot the mass of the largest remnant normalized by the total system mass, $M_\mathrm{{LR}}/M_\mathrm{{tot}}$, as a function of the impact speed normalized by the escape speed from the total system mass (as defined in Section 2.4), $v_{\mathrm{imp}}/v_{\mathrm{esc}}$. Open circles are for the cases where the target has no pre-impact spin, while open triangles are for cases where the target has a pre-impact spin period of 6 hours. As shown in the figures, the mass of the largest remnant is a negative linear function of the impact speed in the gravitational dispersal regime. We find that the energy required to reach the gravitational dispersal regime increases for cases with materials that are more dissipative. In general, we find that including rotation increases the mass loss for a given impact speed systematically, for all material groups tested.  

\begin{center}
\textit{\Fig{mesc_smooth}}.\\
\textit{\Fig{mesc_glass}}.\\
\textit{\Fig{mesc_gravel}}.
\end{center}

\subsection{Characterizing the Dependence on Initial Conditions}

Since collisional dissipation is dependent on the internal configuration of a rubble pile, the collision outcomes might be sensitive to the initial conditions of the rubble-pile target. We performed a set of simulations to determine the dependence of mass loss on the random nature of the initial conditions generation. Five randomly generated rubble-pile targets were constructed through spherical collapse. The equilibrated rubble piles had bulk densities ranging from $1.655$ g cm$^{-3}$ to $1.722$ g cm$^{-3}$. For each material property case, each rubble-pile target was impacted by the nominal projectile with a speed close to catastrophic disruption (6, 10, and 14 m s$^{-1}$ for the corresponding property, respectively). \fig{RandomTest} shows the results of these tests. The rubble piles are ordered from the lowest bulk density (Rubble Pile 1 has a bulk density of 1.655 g cm$^{-3}$) to the highest (Rubble Pile 6 has a bulk density of 1.722 g cm$^{-3}$), reflecting the degree of compaction that was achieved during the collapse phase. The nominal rubble pile used in the previous section is Rubble Pile 5, with a bulk density of 1.705 g cm$^{-3}$. \fig{RandomTest} indicates a weak correlation of mass-loss outcome with target bulk density, with more compacted bodies suffering slightly less mass loss, at the few-percent level. In addition, the dependence of mass-loss outcome with degree of compaction is stronger for the more dissipative material types. Our choice of nominal rubble pile is reasonably representative but a spread of a few percent in mass-loss outcome should be assumed a characteristic uncertainty for the simulations presented here.

\begin{center}
\textit{\Fig{RandomTest}}.
\end{center}

\subsection{Catastrophic Disruption Threshold}
\indent In order to account for the dependence of mass ratio on catastrophic disruption criteria, \citet{LeinhardtStewart09} introduced new variables into their formulation for predicting collision outcomes, the reduced mass $\mu \equiv M_{\mathrm{proj}}M_{\mathrm{targ}}/M_{\mathrm{tot}}$, the reduced-mass specific impact energy $Q_{R}\equiv0.5 \mu v^{2}_{\mathrm{imp}}/M_{\mathrm{tot}}$, and the corresponding reduced-mass catastrophic dispersal limit $Q^{\star}_{RD}$, the specific impact energy required to disperse half the total mass of the system. We use this formulation to describe the catastrophic disruption thresholds of our simulations. For each set of material property and pre-impact rotation values, we determine the catastrophic disruption threshold by performing a linear least-squares fit on the data shown in \figs{mesc_smooth}{mesc_gravel}. We summarize the results of these fits in \tbl{Qstarred}. The errors associated with each value of $Q^{\star}_{RD}$ are the standard deviation of each linear fit. These are found to be small fractions ($\lesssim 2\%$) of the values of $Q^{\star}_{RD}$.\\

\begin{center}
\textit{\tbl{Qstarred}}.
\end{center}

\indent We compare our results to those of \citet{KorycanskyAsphaug09} for the cases of monodisperse rubble piles with a mass ratio of $0.1$. Based on the $Q^{\star}_{RD}$ results, we can compare our three different material properties to their dissipation parameters (normal restitution coefficient, $\epsilon_{n}$, and a friction coefficient, $\eta$, for a Coulomb model of tangential friction). We find $Q^{\star}_{RD}$ values of 1.73, 4.47, and 8.93 J kg$^{-1}$ for the smooth, glass, and gravel cases, respectively. \citet{KorycanskyAsphaug09} find $Q^{\star}_{RD}$ values of 1.87, 3.75, and 8.96 J kg$^{-1}$ for low ($\epsilon_{n}=0.8, \eta=0$), medium ($\epsilon_{n}=0.5, \eta=0$), and high ($\epsilon_{n}=0.5, \eta=0.5$) dissipation parameters, respectively. The parameter values for low dissipation are similar to our smooth case; therefore, we would expect that the values determined for $Q^{\star}_{RD}$ would be similar for the two models. However, at higher dissipation, the coefficient of friction values that the two models use diverge (our work uses much higher values), yet the $Q^{\star}_{RD}$ results are still similar. For these cases, the use of polyhedral particles in \citet{KorycanskyAsphaug09} may contribute to the added shear strength of their rubble piles. Furthermore, they find that these values nearly double if the rubble piles are made up of a power-law size distribution of elements with a power law index of $-1$. Evidently the physical characteristics that determine the shape and spin limits of gravitational aggregates \citep{Walsh12} also contribute to their collisional evolution. The results of our work show that the catastrophic disruption of rubble-pile asteroids is highly dependent on their internal configuration, such that variations in shear strength can contribute to changes of $\sim$ half an order of magnitude in the critical threshold for dispersal. We illustrate this further in \fig{Qstar_ratio}, where we plot $Q^{\star}_{RD}$ normalized by the specific gravitational binding energy, $U$, as a function of each material property. For a binary collision, the gravitational binding energy can be approximated as\\ 
\begin{equation} U=\frac{3GM_{tot}}{5R_{C1}}=\frac{4}{5}\pi\rho_{1}GR^{2}_{C1}, \end{equation}
where $G$ is the gravitational constant and $R_{C1}$ is the spherical radius of the combined projectile and target masses at a density of $\rho_{1}=1$ g cm$^{-3}$. \citet{LeinhardtStewart09} introduced $R_{C1}$ in order to compare collisions of different projectile-to-target-mass ratios.\\
\begin{center}
\textit{\Fig{Qstar_ratio}}.
\end{center}
\indent For cases where the target has pre-impact rotation, $Q^{\star}_{RD}$ systematically decreases by $\sim$ 6\% for each material property. This is consistent with the rotation-dependent catastrophic disruption studies by \citet{Ballouz14}. However, their work focused on a single type of material (i.e., a single set of SSDEM parameters that resemble the glass case). For head-on impacts with impact velocity perpendicular to spin angular momentum, they found that $Q^{\star}_{RD}$ drops by 6--8\% when the target has a pre-impact spin period of 6 hours. We find that this outcome is true for any material property. This result suggests that enhancement of mass loss due to rotation is independent of shear-strength effects. This seems reasonable, since rotation should only influence the effective gravitational binding energy of a gravitational aggregate. Therefore, rotation affects the collisional and shape evolution of gravitational aggregates; however, while the shape evolution is sensitive to the body's internal structure, rotation-enhanced mass loss is not. Furthermore, our results in Section 3.1 show that mass-loss outcomes vary depending on the initial internal configuration of the rubble pile, with the effect being greater for material properties with higher dissipation parameters. Since rotation-induced enhancement in mass loss is independent of the effects of shear strength, it is also likely independent of the initial conditions of the rubble-pile target. Therefore, our determination of the relative mass-loss enhancements for targets with a pre-impact rotation of 6 hours is likely correct for any given rubble-pile target.

\subsection{Mass Loss as a Function of the Critical Threshold for Catastrophic Disruption}
By introducing reduced-mass variables for catastrophic disruption, \citet{LeinhardtStewart09} showed that the outcome of any head-on collision, regardless of projectile-to-target-mass ratio, can be described by a single equation that they call the ``universal'' law for catastrophic disruption: 
\begin{equation}M_{\mathrm{LR}}/M_{\mathrm{tot}} = -0.5 (Q_{R}/Q^{\star}_{RD})+0.5, \end{equation}
In \fig{Universal}, we show that our results are consistent with this formulation. Eq.\ (3) describes well the mass loss outcomes of catastrophic collisions, except for several high-impact-energy cases where the largest remnant retains a higher mass than Eq.\ (3) would suggest. At these energies, a larger fraction of the energy is collisionally dissipated, and several aggregates are able to form with sizes similar to the largest remnant. Therefore, at incrementally higher specific impact energies the mass of a single largest remnant becomes less representative of the collisional dynamics. Correctly using Eq.\ (3) in population models of small-body evolution requires an appropriate choice for $Q^{\star}_{RD}$. As shown in the previous section, depending on the material property used, this value can vary by nearly an order of magnitude.
\begin{center}
\textit{\Fig{Universal}}.
\end{center} 
\section{Conclusions \& Future Work}
Many important aspects of planetary evolution, including the distribution of volatiles in the Solar System and the delivery of water to Earth, are influenced by the collisional and orbital histories of small bodies. How objects fragment in collisions \citep{Leinhardt00, Michel04, Benavidez12}, how the orbits of those fragments spread over time \citep{Bottke13}, and how the fragments behave in tidal interactions, both with planets \citep{HolsappleMichel06, HolsappleMichel08} and with each other \citep{GoldreichSari09}, all are known to be influenced by the degree to which the colliding objects and the fragments are rigid or deformable. Thus, understanding strength is key to understanding the histories of small bodies in general and their role in Solar System evolution. However, very little is known about the internal structures of small bodies. Future space missions, in particular sample-return ones such as Hayabusa 2 (JAXA) to be launched in 2014--2015, and OSIRIS-REx (NASA) to be launched in 2016, will shed some light on the physical and dynamical properties of asteroids that will help constrain the plausible SSDEM values. Until then, we are able to perform simulations that test different types of material responses to account for the wide diversity of physical properties in small-body populations.\\
\indent We find that dissipation and frictional effects combine to increase the catastrophic disruption threshold in rubble-pile collisions. For a range of three different types of material with varying shear strengths, the critical energy required differs by about half an order of magnitude. We also find that pre-impact rotation can decrease the catastrophic disruption threshold independent of a rubble pile's internal configuration. Moreover, the relative change in $Q^{\star}_{RD}$ for cases with pre-impact spin is constant across all the material types studied here.\\
\indent Since our code does not model the fragmentation or deformation of individual particles, we can only explore the effects that strength parameters have on the outcomes of collisions with impact speeds less than the sound speed of the material. Hence, the implications of our results on the hyper-velocity impacts that occur between asteroids today have yet to be tested. At such speeds, the actual fragmentation process damaging the material becomes important. For very porous bodies or for very high impact speeds, compaction and/or heat effects will contribute to energy dissipation \citep[e.g.,][]{Jutzi08}. Thus, the catastrophic disruption criterion is very sensitive to the size of the body, its internal structure, and also the impact speed \citep[e.g.,][]{Jutzi10}.  Furthermore, the effect of pre-impact rotation at such high speeds is not well understood, as the handful of studies that have been performed in this regime have focused on the post-collision spin rate, rather than the mass loss \citep{Canup08,TakedaOhtsuki09}. However, it is reasonable to expect rotation to influence the mass-loss outcomes in collisions between asteroids, whose spin rates can take a large range of values (some are even very fast rotators; e.g., \citeauthor{Holsapple07}, \citeyear{Holsapple07}), even if they occur in the hyper-velocity regime with typical impact speeds about 5 km/s \citep{Bottke94}. In fact, recent hyper-velocity impact experiments indicate mass-loss enhancement effects of rotation, as we find in our low-speed collision modeling \citep{Morris12}.\\
\indent There is an enormous parameter space of rubble-pile collision experiments that remains to be explored. As a first step, in order to describe collisional processes in the Solar System more accurately, the dependence of impact outcomes on the size distribution and cohesional strength between the gravitational aggregate's constituent particles could be explored, which our code can be used to do.

\section*{Acknowledgments}
This material is based on work supported by the U.S.\ National Aeronautics and Space Administration under Grant Nos.\ NNX08AM39G, NNX10AQ01G, and NNX12AG29G issued through the Office of Space Science and by the U.S.\ National Science Foundation under Grant No.\ AST1009579. Part of the research leading to these results received funding from the European Union Seventh Framework Programme (FP7/2007--2013) under grant agreement n\degree 282703--NEOShield. Simulations were performed on the YORP cluster administered by the Center for Theory and Computation, part of the Department of Astronomy at the University of Maryland.

\bibliography{mybib}{}
\bibliographystyle{elsarticle-harv}

%
%Tables
%
\newpage
\renewcommand{\baselinestretch}{1} % tables are to be single-spaced
\begin{table}[h]
  \centering
  \captionsetup{format=plain, justification=centering}
  \caption{\\Summary of Material SSDEM Parameters}
  \label{t:SSDEMparams}
  \begin{tabular}{c|ccc}
    Parameters & Smooth & Glass Beads & Gravel\\
    \hline
    $\mu_s$ & 0.0 & 0.43 & 1.31 \\
    $\mu_r$ & 0.0 & 0.1 & 3.0 \\
    $\varepsilon_n$ & 0.95 & 0.95 & 0.55 \\
    $\varepsilon_t$ & 1.0 & 1.0 & 0.55 \\
    \hline
    \hline
  \end{tabular}
\end{table}

\newpage
\renewcommand{\baselinestretch}{1} % tables are to be single-spaced
\begin{table}[h]
  \centering
  \captionsetup{format=plain, justification=centering}
  \caption{\\Summary of Catastrophic Disruption Thresholds}
  \label{t:Qstarred}
  \begin{tabular}{c|cc}
 Material & $P_\mathrm{{spin}}$ (h) & $Q^{\star}_{RD}$ (J kg$^{-1}$) \\ 
\hline
smooth & $\infty$ & 1.73 $\pm$ 0.03 \\ 
smooth & 6 &  1.60 $\pm$ 0.03\\ 
glass & $\infty$ & 4.47 $\pm$ 0.01 \\ 
glass & 6 &  4.24 $\pm$ 0.02\\ 
gravel & $\infty$ & 8.93 $\pm$ 0.09   \\ 
gravel & 6 & 8.36 $\pm$ 0.04  \\ 
\hline
\hline
  \end{tabular}
\end{table}

%
%Figure Captions
%
\newpage
\section*{Figure Captions}

\begin{figure}[h!]\centering\includegraphics{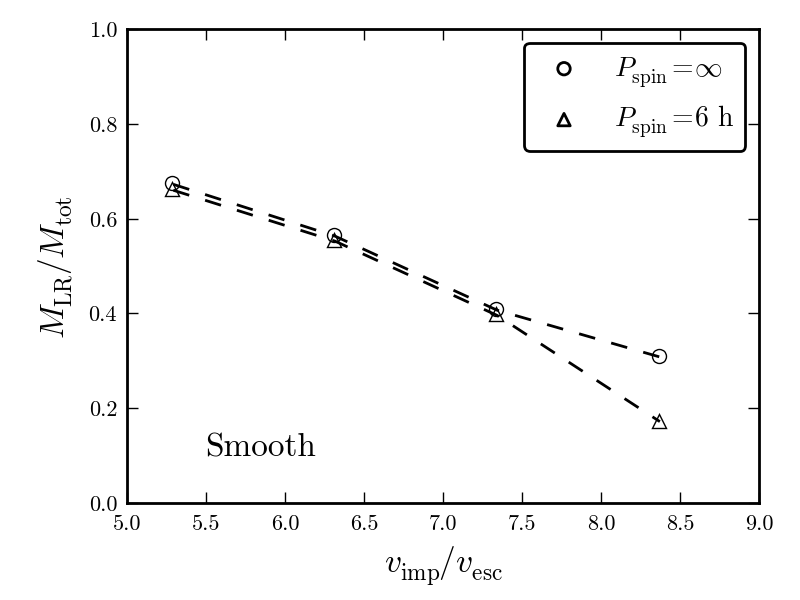}\caption{}\label{f:mesc_smooth}\end{figure}

\begin{figure}[h!]\centering\includegraphics{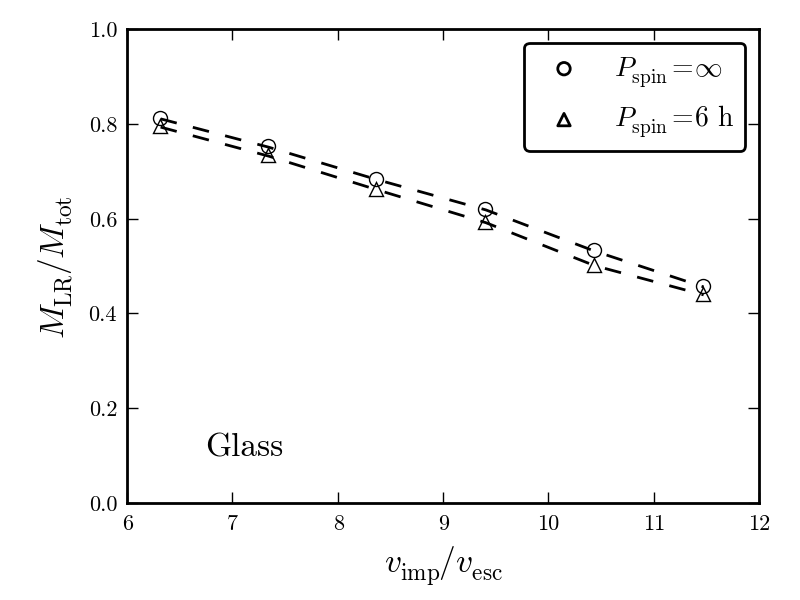}\caption{}\label{f:mesc_glass}\end{figure}

\begin{figure}[h!]\centering\includegraphics{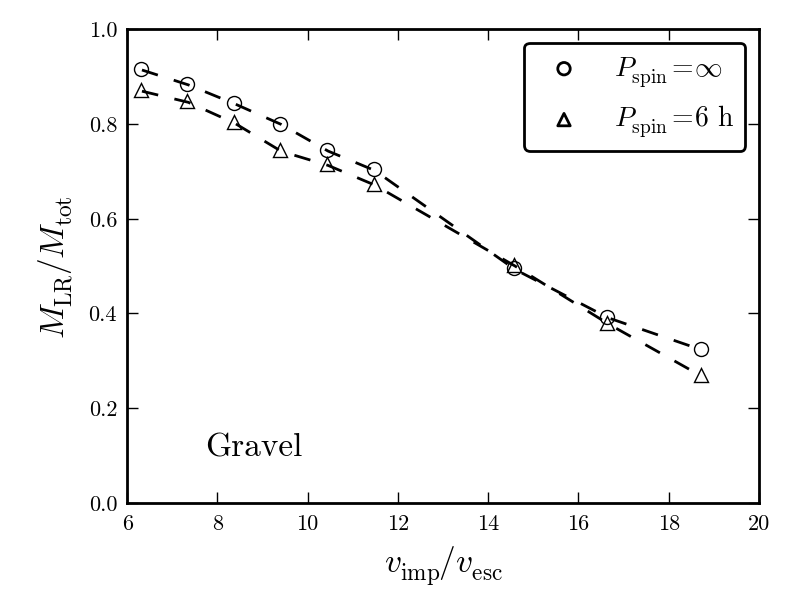}\caption{}\label{f:mesc_gravel}\end{figure}

\begin{figure}[h!]\centering\includegraphics{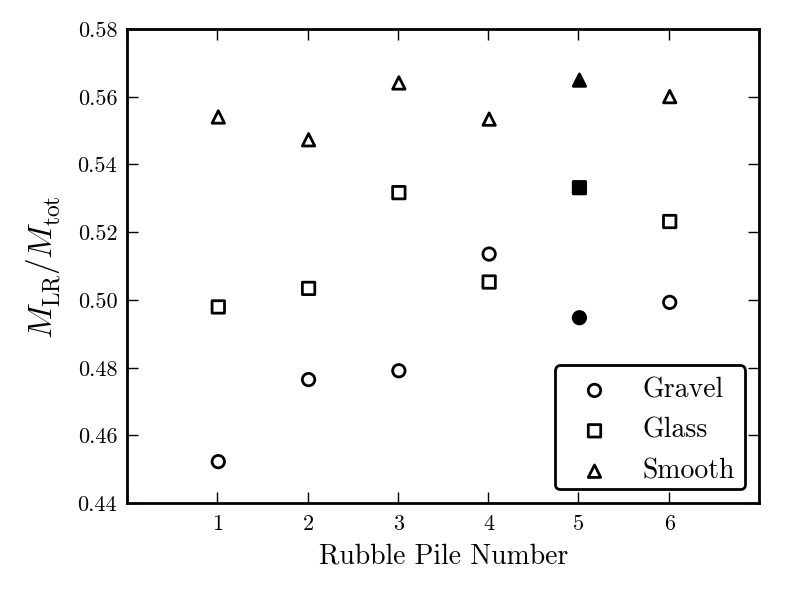}\caption{}\label{f:RandomTest}\end{figure}

\begin{figure}[h!]\centering\includegraphics{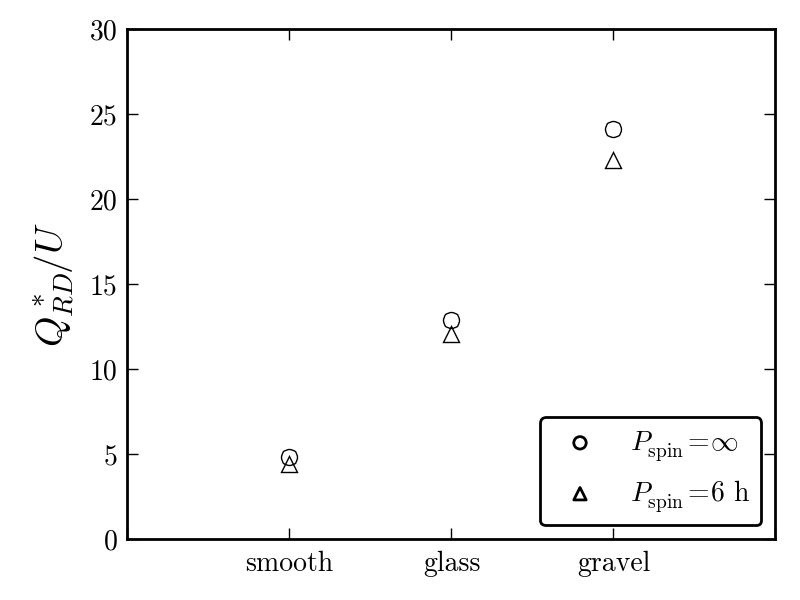}\caption{}\label{f:Qstar_ratio}\end{figure}

\begin{figure}[h!]\centering\includegraphics{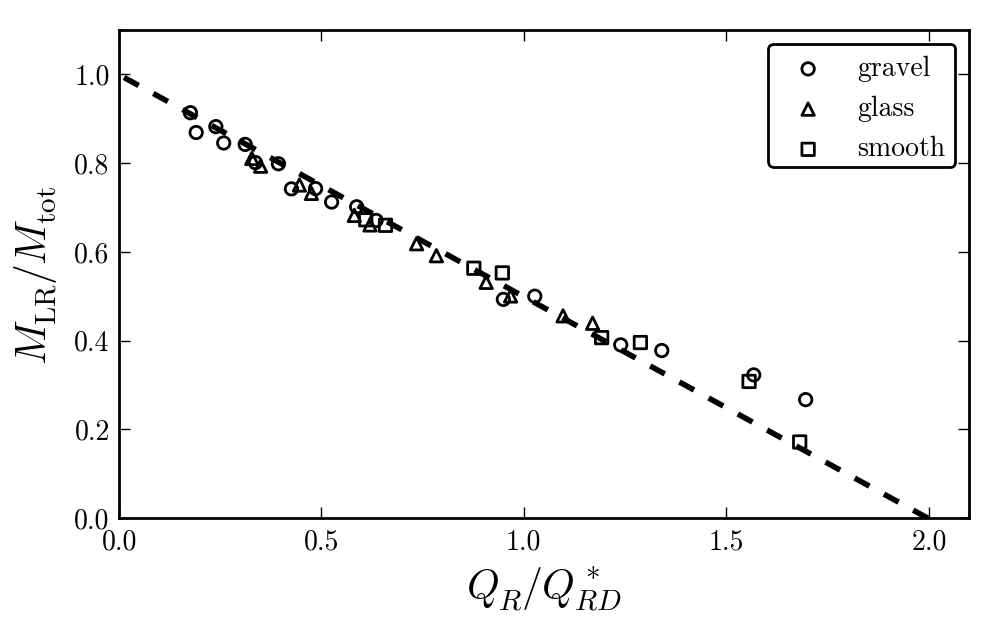}\caption{}\label{f:Universal}\end{figure}

\begin{description}

\figcap{mesc_smooth}{The mass of the largest remnant for the ``smooth'' material set of collisions. Open circles are for collisions with no pre-impact spin. Open triangles are for collisions where the target has a pre-impact spin of 6 hours. Spin systematically increases the amount of mass lost for a collision. This effect is enhanced at high impact speeds.}

\figcap{mesc_glass}{The mass of the largest remnant for the ``glass'' material set of collisions. Compared to Fig. 1 for the "smooth" case, the largest remnant retains more mass at all impact speeds (note the difference in scale on the horizontal axis).  The trend of increased mass loss at higher spin persists.}

\figcap{mesc_gravel}{The mass of the largest remnant for the ``gravel'' material set of collisions. Compared to the "smooth" and "glass" cases (Figs.\ 1 \& 2), even more mass is retained at a given impact speed, while, in general, spin continues to enhance mass loss (the effect is diminished in this case close to catastrophic disruption.)}

\figcap{RandomTest}{For different randomizations of the rubble-pile target, the mass of the largest remnant varies by roughly 1\%, 2\%, and 3\% for the smooth (triangles), glass (squares), and gravel (circles) cases, respectively. The filled-in symbols represent the nominal rubble-pile targets used to determine the dependence of mass loss with impact speed and pre-impact rotation for each material property tested.}

\figcap{Qstar_ratio}{The reduced-mass catastrophic disruption threshold, $Q^{\star}_{RD}$, normalized by the gravitational binding energy, $U$, for the three material properties studied here. Open circles are for collisions with no pre-impact spin. Open triangles are for collisions where the target has a pre-impact spin of 6 hours. Spin systematically decreases $Q^{\star}_{RD}/U$ by $\sim$ 6\%.}

\figcap{Universal}{Except for three impacts at high energies, collision outcomes are described well ($< 10\%$ difference) by the ``universal'' law for catastrophic disruption, Eq.\ (3) (black dotted-line).}

\end{description}

%
%Figures
%
\newpage

\end{document}